# Impact of Electrode Position on Forearm Orientation Invariant Hand Gesture Recognition

Md. Johirul Islam, Umme Rumman, Arifa Ferdousi, Md. Sarwar Pervez, Iffat Ara, Shamim Ahmad, *Member, IEEE*, Fahmida Haque, Sawal Hamid, Md. Ali, *Senior Member, IEEE*, Kh Shahriya Zaman, *Member, IEEE*, Mamun Bin Ibne Reaz*, *Senior Member, IEEE*, Mustafa Habib Chowdhury* *Senior Member, IEEE,* Md. Rezaul Islam

*Abstract*— **Objective: Variation of forearm orientation is one of the crucial factors that drastically degrades the forearm orientation invariant hand gesture recognition performance or the degree of freedom and limits the successful commercialization of myoelectric prosthetic hand or electromyogram (EMG) signal-based human-computer interfacing devices. This study investigates the impact of surface EMG electrode positions (elbow and forearm) on forearm orientation invariant hand gesture recognition. *Methods:* The study has been performed over 19 intact limbed subjects, considering 12 daily living hand gestures. The quality of the EMG signal is confirmed in terms of three indices. Then, the recognition performance is evaluated and validated by considering three training strategies, six feature extraction methods, and three classifiers. *Results:* The forearm electrode position provides comparable to or better EMG signal quality considering three indices. In this research, the forearm electrode position achieves up to 5.35% improved forearm orientation invariant hand gesture recognition performance compared to the elbow electrode position. The obtained performance is validated by considering six feature extraction methods, three classifiers, and real-time experiments. In addition, the forearm electrode position shows its robustness with the existence of recent works, considering recognition performance, investigated gestures, the number of channels, the dimensionality of feature space, and the number of subjects. *Conclusion:* The forearm electrode position can be the best choice for getting improved forearm orientation invariant hand gesture recognition performance. *Significance:* The performance of myoelectric prosthesis and human-computer interfacing devices can be improved with this optimized electrode position.**

*Index Terms*— **Forearm orientation generalization, hand gesture recognition, electrode position optimization, and forearm rotation.**

This work was financially supported by the Ministry of Science and Technology, Bangladesh under reference grant code: SRG-232388, Vice Chancellor's Research Fund of Independent University, Bangladesh under reference grant code: VCRF-SETS:24-021, and Rajshahi University of Engineering & Technology (RUET) under grant no DRE/7/RUET/574(58)/PRO/2023-24/50.

*Corresponding Authors: Mamun Bin Ibne Reaz and Mustafa Habib Chowdhury*

Md. Johirul Islam and Md. Sarwar Pervez are with the Department of Physics, Rajshahi University of Engineering and Technology, Rajshahi-6204, Bangladesh (johirul@phy.ruet.ac.bd, sarwar@phy.ruet.ac.bd).

Umme Rumman is with the Department of Computer Science and Engineering, Varendra University, Rajshahi, Bangladesh. (chaitycse9@gmail.com)

Arifa Ferdousi, and Sawal Hamid Md. Ali are with the Department of Electrical, Electronic and Systems Engineering, Universiti of Kebangsaan, Malaysia, Bangi, Selangor, Malaysia. (afpris@gmail.com, sawal@ukm.edu.my).

Iffat Ara and Md. Rezaul Islam is with the Department of Electrical and Electronic Engineering, University of Rajshahi, Rajshahi-6205, Bangladesh. (iffat.ice@pust.ac.bd, rima@ru.ac.bd)

Shamim Ahmad is with the Department of Computer Science and Engineering, University of Rajshahi, Rajshahi-6205, Bangladesh. (shamim_cst@ru.ac.bd).

Fahmida Haque is with Artificial Intelligence Resource, Molecular Imaging Branch, National Cancer Institute, Bethesda, MD, USA. (fahmida.haque@nih.gov).

Kh Shahriya Zaman, Mamun Bin Ibne Reaz, and Mustafa Habib Chowdhury are with the Department of Electrical and Electronic Engineering, Independent University, Bangladesh; Dhaka, Bangladesh. (shahriya@iub.edu.bd, mamun.reaz@iub.edu.bd, mchowdhury@iub.edu.bd)

## I. INTRODUCTION

THE EMG signal is generated in the muscle during performing any gestures that are controlled by the central nervous system or artificial excitation. So, the EMG signal carries information to the respective gesture. The EMG signal can be acquired through the needle, surface, or recently proposed capacitive electrodes [1], [2]. Among the three techniques, surface electromyography is popular for its risk-free acquisition and simplicity. The EMG signal is relatively high in amplitude (mV) compared to the electroencephalogram, having a high bandwidth of 20 to 500 Hz [3]. Then, the EMG signal is often chosen for hand gesture recognition [4], [5], which is commonly employed for myoelectric prosthetic hand [6], and human-computer interfacing [7]. However, myoelectric pattern recognition-based devices could not go for massive production due to their few degrees of freedom and limited hand gesture recognition performance, highly influenced by forearm orientation [8], [9], muscle contraction force [10], electrode shift [11], mobility [12], multi-day variation [13], and cross-user variability [14]. In a real-time scenario, multiple factors are introduced simultaneously which drastically degrade the recognition performance. Then, the degree of freedom is compromised to optimize the recognition performance [15], [16]. Sometimes, the number of EMG channels is increased up to 128 or 196 (high-density electrode array), but a large number of electrodes increases system complexity, user unfriendliness, and computational power as well [17], [18]. In addition, the optimization of electrode position [19], windowing strategy [20], feature extraction [21], [22], feature selection [23], [24], dimension reduction [25], [26], and machine learning and deep-learning algorithms [27], [28], [29], are employed to improve the recognition performance with performance degrading factors. However, algorithm-based solutions often require high computational power and are not generalized to address multiple factors [15], [25].

One of the dominant performance degrading factors in myoelectric pattern recognition is forearm orientation change where the responsible muscle (pronator teres, pronator quadratus, and supinator) is activated additionally with the hand gesture performing muscles, resulting in changes in the EMG signal due to the superposition. Additionally, the rotation of radio and ulna bones during any forearm orientation shifts the attached muscles relative to the surface electrodes on the skin. However, recent studies focused on recognition algorithms rather than finding the background reasons. Fougner *et al.* [30] reported that limb position and its orientation can significantly affect hand gesture recognition performance. They employed an accelerometer sensor additionally with the EMG electrodes and showed that it could improve the recognition performance by up to 13%. Khushaba *et al.* [31] also utilized the concept of an accelerometer to solve both factors of the forearm orientation and muscle contraction force. To recognize six hand gestures, they employed six feature extraction methods with a support vector machine (SVM) as a classifier. They observed that time-domain power spectral descriptors proposed by Khushaba *et al.* [32] and Al-Timemy *et al.* [33], achieved the highest recognition performance of about 91% in muscle contraction force invariant hand gesture recognition and about 60.67% in forearm orientation invariant hand gesture recognition, individually. Therefore, unknown muscle contraction force was successfully recognized by their recommended feature extraction method, however, unknown forearm orientations were not recognized properly. Further, Rajapriya *et al.* [34] proposed a feature extraction method based on wavelet bispectrum to improve the forearm invariant hand gesture recognition performance and achieved 86.43% recognition performance using EMG signals only. Thereafter, Islam *et al.* [8] proposed an EMG channel and feature selection method and achieved 78.44% of both forearm orientation and muscle force invariant hand gesture recognition. They also employed pronation and supination with medium muscle contraction force for training and achieved an improved recognition performance of 91.46%. Therefore, the feature extraction method can improve the forearm orientation invariant hand gesture recognition performance, but a successful forearm invariant hand gesture recognition requires an alternative solution that is still unsolved by researchers.

Botros *et al.* [35] investigated forearm (near elbow) and wrist electrode positions to find an optimum electrode position for hand gesture recognition and noticed that the wrist electrode position significantly performed better in terms of the EMG signal quality and recognition performance as well. In the day-to-day stability of the EMG signal, the wrist electrode position also performed better [36]. The best-performed wrist position is available for the intact limed subject, but very challenging to avail for an amputee. To address the issue, Islam *et al.* [19] investigated eight 1D electrode arrays, starting from the elbow to the middle of the forearm. They noticed that the electrode array placed between the elbow and wrist (on the middle of the forearm) achieved significantly higher hand gesture recognition performance. Therefore, an optimized electrode position can be helpful in improving the hand gesture recognition performance. Then, it is expected that the forearm orientation can be minimized by finding an optimized electrode position on the forearm. However, to the best of our knowledge, the impact of electrode position was not investigated by anyone, considering multiple electrode positions.

In this context, the most available electrode positions (elbow and forearm) for amputee and intact limbed subjects have been investigated to improve forearm invariant hand gesture recognition performance. In this study, the highest number of subjects are incorporated compared to related studies. First, the quality of the EMG signal is confirmed in terms of three indices. Then, the recognition performance is evaluated and validated considering six feature extraction methods, and three classifiers. In this research, multiple training strategies are employed to achieve a satisfactory forearm invariant hand gesture recognition performance. All the best-performing cases are confirmed by a statistical test. In addition, the impact of the the electrode positions is confirmed in real-time experiments. Thereafter, the obtained performances are also compared and validated by existing works. Finally, multiple recommendations with some future research studies have been proposed.

## II. METHODOLOGY

### A. Signal Acquisition

A total of 19 healthy subjects (16 males and 3 females) voluntarily participated in EMG data acquisition, aged between 25 to 40 years. The participants were informed about the detailed procedure of the acquisition and the purpose of the experiment. Then, written consent was taken from each participant. Ethical approval was taken from the Dean of Applied Science and Humanities, Rajshahi University of Engineering & Technology. In this acquisition, a customized EMG signal acquisition system was employed [37], which facilitated eight-channel EMG signal acquisition with a sampling frequency of 1000 Hz at a resolution of 10 bits and SNR value of 35 dB depending upon the gesture and muscle contraction force. Four active electrodes denoted CH1 to CH4 (modified from the MFI bar electrode, MFI Medical Equipment, Inc., USA) were placed around the elbow, and another four active electrodes denoted CH5 to CH8 were placed around the forearm where their common ground electrode (MFI Medical Equipment, Inc., USA) was placed on the dorsal side of the hand (Fig. 1a). The electrode positions with underlying muscles and their functions are noted in Table I. In this study, each of the subjects performed 12 daily living hand gestures, considering single-digit, multiple-digit, and wrist gestures. The gestures included thumb up (TU), index (IDX), right angle (RA), peace (PCE), index little (IL), thumb little (TL), hand close (HC), hand open (HO), wrist extension (WE), wrist flexion (WF), ulnar deviation (UD), and radial deviation (RD). The subject performed each gesture five times (trials) with a duration of 8 seconds, starting from a relaxed position and maintaining a resting time of 1 minute between two successive trials. The subject repeated 12 gestures for three forearm orientations (pronation, rest, and supination) shown in Fig. 1b. Therefore, the total number of signals collected from each subject was 180 (180×8 seconds = 1440 seconds).

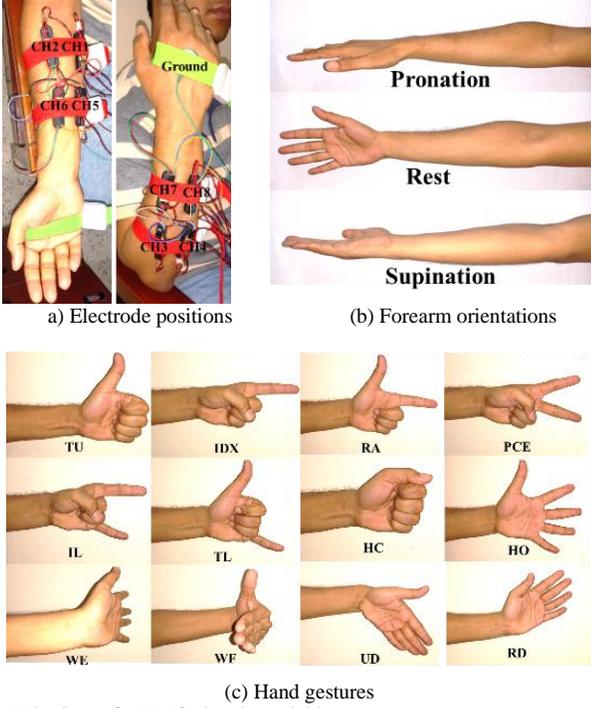

(a) Electrode positions  (b) Forearm orientations

(c) Hand gestures
Fig. 1 The Setup for EMG signal acquisition.

TABLE I: ELECTRODE POSITIONS WITH UNDERLYING MUSCLES AND THEIR FUNCTIONS.

| | Underlying Muscle | Function |
|---|---|---|
| CH1 | Flexor carpi radialis | Wrist flexion and abduction |
| | Palmoris longus | Wrist Flexion |
| | Flexor digitorum superficialis | Flexion of the index to little fingers |
| | Flexor digitorum profundus | Flexion of the index to little fingers |
| | Pronator teres | Pronation |
| CH2 | Brachioradialis | Weak elbow flexion |
| | Edge of the flexor digitorum superficialis | Flexion of the index to little fingers |
| | Edge of the flexor digitorum profundus | Flexion of the index to little fingers |
| CH3 | Extensor digitorum communis | Extension of the index to little fingers |
| | Extensor carpi radialis brevis | Abduction and wrist extension |
| | Extensor digiti minimi | Extension of little finger |
| CH4 | Extensor carpi ulnaris | Adduction and extension of the wrist |
| | Flexor carpi ulnaris | Adduction |
| CH5 | Flexor carpi radialis | Wrist flexion and abduction |
| | Palmoris longus | Wrist Flexion |
| | Flexor digitorum superficialis | Flexion of the index to little fingers |
| | Flexor digitorum profundus | Flexion of the index to little fingers |
| | Pronator teres | Pronation |
| CH6 | Brachioradialis | Weak elbow flexion |
| | Edge of the flexor digitorum superficialis | Flexion of the index to little fingers |
| | Flexor pollicis longus | Flexion of the thumb |
| CH7 | Extensor digitorum communis | Extension of the index to little fingers |
| | Extensor carpi radialis brevis | Abduction and wrist extension |
| | Extensor digiti minimi | Extension of little finger |
| CH8 | Extensor carpi ulnaris | Adduction and extension of the wrist |
| | Flexor carpi ulnaris | Adduction |

*B. Signal Quality Evaluating Indices*

The EMG signal quality can be evaluated using three indices including SNR, SMR, and FER which can be employed to compare elbow and forearm electrode positions [35], [19]. The SNR value indicates the EMG signal strength compared to the noise produced during resting conditions. The SNR value is evaluated in dB according to (1) where a higher value indicates stronger signal strength. Again, the SMR value shows the EMG signal strength to the motion artifact which is almost unavoidable noise generated due to the motion of the electrode during muscle contraction. So, the EMG signal having a lower motion artifact indicated by higher SMR is considered a high-quality signal. It can be evaluated using (2). In addition, the EMG signal strength of two electrode positions can be compared employing the FER value and evaluated employing (3). The higher FER value (>1) indicates the forearm signal strength is higher than the elbow signal strength or vice versa.

$$SNR = 20\log_{10}\left[\frac{\sqrt{Raw\ EMG\ Signal_{RMS}^2 - Noise_{RMS}^2}}{Noise_{RMS}}\right] \quad (1)$$

$$SMR = 10\log\left[\frac{\sum_{f=20}^{500} PSD}{\sum_{f=0}^{20} PSD}\right] \quad (2)$$

$$FER = \frac{\left[\sqrt{Raw\ EMG\ Signal_{RMS}^2 - Noise_{RMS}^2}\right]_{Forearm}}{\left[\sqrt{Raw\ EMG\ Signal_{RMS}^2 - Noise_{RMS}^2}\right]_{Elbow}} \quad (3)$$

*C. Forearm Orientation Invariant Hand Gesture Recognition*

In this study, forearm orientation invariant hand gesture recognition performance was evaluated employing MATLAB 2020a (MathWorks, USA). First, the active EMG signal was segmented from 8 seconds of recorded signal. The length of the active signal varied between 5 to 7 seconds depending upon the hand gesture activation from the resting condition. To minimize motion artifacts and high-frequency noise, the raw EMG signal was preprocessed employing a bandpass filter of 20 to 450 Hz [38]. In addition, a notch filter of 50 Hz was employed to suppress power line artifacts [37]. Further, the active EMG signal was segmented employing disjoint windowing of 250 ms, so that the total processing time lies within the recommended delay of 300 ms [39]. In this research, the features were extracted employing six reputed EMG feature extraction methods, including signal normalized time domain features (SNTDF) [3], Hjorth secant line features (HSL) [40], Temporal-spatial descriptors (TSD) [21], time-domain descriptors (TDD) [41], fusion-based time-domain descriptors (FTDD) [32], six-order autoregressive coefficients with root mean square value (AR-RMS) [42]. The SNTDF, HSL, TSD, TDD, FTDD, and AR-RMS include seven features and correlation coefficients for each pair of channels, five features based on Hjorth parameters, seven features extracted from each channel and each pair of the difference of channels, five features based on indirect Fourier transform properties and sum of squares differences, six features based on the orientation between the original features and a nonlinearly transformed features of them, and six-order autoregression coefficients with RMS value, respectively. Therefore, for each array (elbow and forearm), the dimensions of the feature space of SNTDF, HSL, TSD, TDD, FTDD, and AR-RMS were 34, 20, 49, 20, 24, and 28, respectively. The higher dimension was reduced to 11 using spectral regression discriminant analysis [43]. In forearm orientation invariant hand gesture recognition, a classifer is trained with the hand gestures of known orientations but tested with the hand gestures of known orientations and unknown orientations as well. To evaluate the performance of forearm orientation invariant hand gesture recognition, a 5-fold cross-validation technique was employed and performed specially so that testing orientations do not mix with the training orientations [8]. First, samples of known and unknown orientations were split into training and testing samples. Then, the model is trained with the samples of known orientations only but tested with the samples of both known and unknown orientations. In this research, three well-performed and simple classifiers, namely k-nearest neighbors (KNN) with Euclidean distance and ten neighbors, support vector machine (SVM) with a Gaussian radial basis optimization (kernel scale=3), and linear discriminant analysis (LDA) were employed and tuned using '*Classification Learner*' app recommended in [4], [40]. Finally, the performance was evaluated using six statistical parameters, including accuracy, sensitivity, specificity, precision, F1 score, and Matthews correlation coefficient (MCC) [8]. These parameters are defined as follows:

$$Accuracy = \frac{TP+TN}{TP+TN+FP+FN} \quad (4)$$

$$Sensitivity = \frac{TP}{TP+FN} \quad (5)$$

$$Specificity = \frac{TN}{TN+FP} \quad (6)$$

$$Precision = \frac{TP}{TP+FP} \quad (7)$$

$$F1\ Score = \frac{2 \times Precision \times Sensitivity}{Precision + Sensitivity} \quad (8)$$

$$MCC = \frac{TN \times TP - FN \times FP}{\sqrt{(TP+FP)(TP+FN)(TN+FP)(TN+FN)}} \quad (9)$$

Where *TP, TN, FP,* and *FN* denote the number of the true positive hand gestures, the true negative hand gestures, the false positive hand gestures, and the false negative hand gestures, respectively.

### D. Statistical Test

To find the significant difference between elbow and forearm electrode positions, analysis of variance (ANOVA) was employed, considering SNR and SMR of 12 hand gestures. In addition, three-way ANOVA was performed to compare the F1 score, considering electrode positions, feature extraction methods, and classifiers in Sections 3.2 and 3.3. Also, a two-way ANOVA was performed to compare the F1 score, considering feature extraction methods and classifiers in Section 3.4. In all cases, a threshold of 5% was considered, and the results were Bonferroni corrected. The obtained *p*-values below 0.05 indicated that the indices or F1 scores were statistically significant.

## III. RESULTS

### A. The EMG Signal Quality

TABLE II
THE COMPARISON OF THE EMG SIGNAL QUALITY BETWEEN ELBOW AND FOREARM ELECTRODE POSITIONS IN TERMS OF SNR, SMR, AND FER.

|    | SNR (dB) Elbow | SNR (dB) Forearm | p | SMR Elbow | SMR Forearm | p | FER |
|----|------|---------|------|---------|---------|------|---------|
| TU | 28.53±11.0 | 30.03±9.96 | 0.52 | 8.22±0.9 | **8.75±2.08** | 0.26 | 1.33±0.50 |
| ID | 23.85±8.98 | 27.34±10.0 | 0.15 | 7.51±1.3 | **7.66±2.20** | 0.80 | 1.57±0.70 |
| RA | 25.52±8.18 | ***30.66±10.6*** | * | 7.56±1.0 | **7.97±1.86** | 0.35 | 1.60±0.68 |
| PC | 23.86±7.36 | 19.52±10.5 | 0.05 | 7.84±1.1 | **8.47±1.90** | 0.12 | 0.93±0.42 |
| IL | 22.38±9.05 | ***27.67±9.17*** | * | 7.94±1.0 | **8.21±1.54** | 0.42 | 1.54±0.61 |
| TL | 24.88±9.62 | ***30.64±9.04*** | * | 7.86±0.9 | **8.02±1.86** | 0.65 | 1.55±0.48 |
| HC | 33.89±9.39 | 31.46±10.4 | 0.17 | 8.17±0.8 | **8.15±1.45** | 0.97 | 1.14±0.34 |
| HO | 28.02±9.47 | 25.38±9.34 | 0.12 | 8.76±1.0 | **8.76±1.73** | 1.00 | 1.06±0.35 |
| WE | 35.17±7.82 | 28.14±12.5 | * | 9.56±1.0 | **10.12±1.9** | 0.19 | 0.91±0.31 |
| WF | 29.08±7.07 | 25.21±4.87 | * | 9.79±1.5 | **9.03±1.82** | 0.06 | 0.98±0.27 |
| UD | 32.06±6.04 | 26.72±7.18 | * | 10.24±1. | 8.99±1.44 | * | 0.93±0.21 |
| RD | 25.24±11.0 | 22.79±9.77 | 0.11 | 8.07±1.3 | ***9.17±1.80*** | * | 0.95±0.41 |

*Note: \* indicates the significant difference when the p-value is less than 0.05. Italic and bold font indicates the value on the forearm electrode position is significantly higher or comparable with the elbow electrode position.*

The EMG signal quality of elbow and forearm electrode positions was evaluated and compared, considering SNR, SMR, and FER. The average value of the indices for each hand gesture including their respective standard deviation among the subjects and p-values are presented in Table II. The experimental results indicated that the SNR value of elbow and forearm electrode positions varied between 22.38 to 35.17 dB and 19.52 to 31.46 dB, respectively. However, in individual hand gesture-wise comparison, the SNR value of forearm electrode position is significantly higher for three hand gestures (RA, IL, TL) or comparable for six hand gestures (TU, IDX, PCE, HC, HO, and RD). Following the trend of SNR value, the SMR value of elbow and forearm electrode position was consistent and varied between 7.51 to 10.24, and 7.66 to 10.12, respectively. In individual comparison, the SMR value of forearm electrode position was higher for one hand gesture (RD) and comparable for ten hand gestures (TU, IDX, RA, PCE, IL, TL, HC, HO, WE, WF). In this study, the FER indices depicted a higher strength of the EMG signal on the forearm electrode position compared to the elbow electrode position for seven hand gestures (TU, IDX, RA, IL, TL, HC, HO). In these gestures, the FER value was within the range of 1.06 to 1.60. Therefore, the forearm electrode position demonstrated an improved or consistent EMG signal quality in terms of SNR, SMR, and FER.

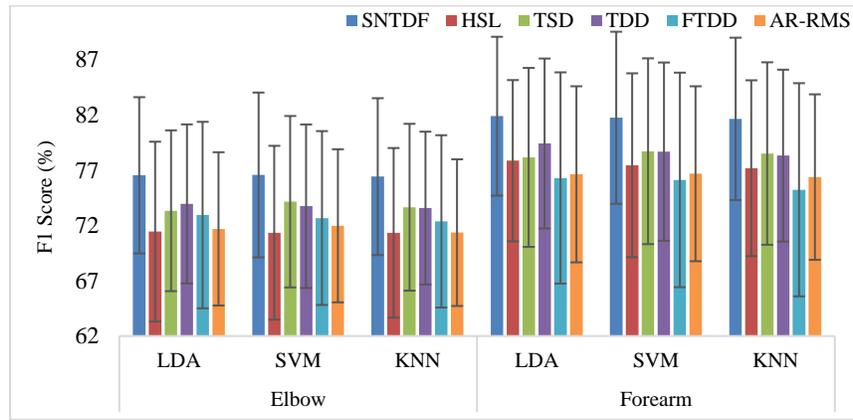

Fig. 2. Forearm orientation invariant hand gesture recognition performance with electrode positions, feature extraction methods, and classifiers.

TABLE III
FOREARM ORIENTATION INVARIANT HAND GESTURE RECOGNITION PERFORMANCES WITH ELECTRODE POSITIONS, CLASSIFIERS, AND FEATURE EXTRACTION METHODS

|  |  |  | SNTDF | HSL | TSD | TDD | FTDD | AR-RMS |
|---|---|---|---|---|---|---|---|---|
| **Elbow** | Accuracy | LDA | 96.08±1.24 | 95.26±1.40 | 95.53±1.29 | 95.66±1.26 | 95.49±1.46 | 95.26±1.20 |
|  |  | SVM | 96.12±1.26 | 95.27±1.34 | 95.71±1.33 | 95.66±1.27 | 95.49±1.33 | 95.36±1.17 |
|  |  | KNN | 96.05±1.25 | 95.24±1.34 | 95.57±1.35 | 95.59±1.22 | 95.40±1.34 | 95.21±1.14 |
|  | Sensitivity | LDA | 76.50±7.43 | 71.55±8.42 | 73.21±7.75 | 73.95±7.54 | 72.94±8.76 | 71.54±7.22 |
|  |  | SVM | 76.72±7.58 | 71.61±8.02 | 74.23±7.95 | 73.93±7.59 | 72.95±7.98 | 72.19±7.00 |
|  |  | KNN | 76.28±7.51 | 71.46±8.01 | 73.42±8.09 | 73.57±7.30 | 72.42±8.02 | 71.27±6.86 |
|  | Specificity | LDA | 97.87±0.68 | 97.42±0.78 | 97.57±0.71 | 97.64±0.69 | 97.54±0.80 | 97.42±0.66 |
|  |  | SVM | 97.89±0.69 | 97.43±0.73 | 97.67±0.72 | 97.64±0.69 | 97.54±0.72 | 97.48±0.64 |
|  |  | KNN | 97.85±0.69 | 97.42±0.73 | 97.59±0.73 | 97.61±0.66 | 97.49±0.73 | 97.40±0.62 |
|  | Precision | LDA | 79.52±5.99 | 75.28±6.9 | 76.93±6.45 | 77.47±6.00 | 76.65±7.17 | 74.68±6.24 |
|  |  | SVM | 78.98±6.37 | 74.73±6.97 | 76.99±6.94 | 76.87±6.13 | 75.79±6.83 | 74.48±6.32 |
|  |  | KNN | 79.14±6.02 | 74.91±6.48 | 77.05±6.51 | 76.77±5.58 | 75.56±6.61 | 74.06±5.99 |
|  | F1 Score | LDA | 76.55±7.07 | 71.45±8.14 | 73.34±7.28 | 73.96±7.20 | 72.95±8.44 | 71.70±6.94 |
|  |  | SVM | 76.58±7.46 | 71.35±7.87 | 74.16±7.76 | 73.75±7.40 | 72.68±7.87 | 71.98±6.93 |
|  |  | KNN | 76.43±7.10 | 71.34±7.68 | 73.66±7.55 | 73.58±6.92 | 72.38±7.80 | 71.36±6.64 |
|  | MCC | LDA | 0.75±0.07 | 0.70±0.09 | 0.72±0.08 | 0.73±0.08 | 0.72±0.09 | 0.70±0.07 |
|  |  | SVM | 0.75±0.08 | 0.70±0.08 | 0.73±0.08 | 0.72±0.08 | 0.71±0.08 | 0.70±0.07 |
|  |  | KNN | 0.75±0.08 | 0.70±0.08 | 0.72±0.08 | 0.72±0.07 | 0.71±0.08 | 0.70±0.07 |
| **Forearm** | Accuracy | LDA | 97.01±1.20 | 96.32±1.23 | 96.38±1.36 | 96.59±1.29 | 96.05±1.65 | 96.09±1.38 |
|  |  | SVM | 97.00±1.30 | 96.26±1.40 | 96.48±1.40 | 96.48±1.34 | 96.01±1.70 | 96.13±1.35 |
|  |  | KNN | 96.97±1.22 | 96.21±1.34 | 96.44±1.38 | 96.41±1.30 | 95.85±1.69 | 96.06±1.28 |
|  | Sensitivity | LDA | 82.05±7.20 | 77.94±7.39 | 78.28±8.16 | 79.56±7.73 | 76.29±9.93 | 76.53±8.27 |
|  |  | SVM | 81.98±7.78 | 77.57±8.40 | 78.91±8.42 | 78.87±8.07 | 76.07±10.2 | 76.79±8.08 |
|  |  | KNN | 81.84±7.34 | 77.26±8.05 | 78.65±8.28 | 78.48±7.79 | 75.12±10.16 | 76.38±7.67 |
|  | Specificity | LDA | 98.38±0.65 | 98.00±0.67 | 98.03±0.74 | 98.15±0.71 | 97.85±0.90 | 97.88±0.74 |
|  |  | SVM | 98.37±0.70 | 97.96±0.76 | 98.08±0.76 | 98.08±0.73 | 97.83±0.92 | 97.90±0.73 |
|  |  | KNN | 98.35±0.67 | 97.93±0.73 | 98.06±0.75 | 98.05±0.71 | 97.75±0.92 | 97.86±0.69 |
|  | Precision | LDA | 84.17±6.12 | 80.41±6.32 | 80.68±6.9 | 81.76±6.81 | 79.62±8.02 | 79.37±6.97 |
|  |  | SVM | 84.01±6.43 | 80.07±7.18 | 81.16±7.11 | 81.33±6.89 | 79.52±8.21 | 79.29±6.81 |
|  |  | KNN | 83.65±6.37 | 79.66±7.00 | 80.8±7.16 | 80.65±6.85 | 78.62±8.01 | 78.89±6.51 |
|  | F1 Score | LDA | 81.90±7.19 | 77.87±7.30 | 78.17±8.10 | 79.43±7.69 | 76.31±9.56 | 76.64±7.97 |
|  |  | SVM | 81.76±7.79 | 77.46±8.32 | 78.73±8.41 | 78.69±8.07 | 76.13±9.71 | 76.69±7.92 |
|  |  | KNN | 81.66±7.35 | 77.19±7.96 | 78.52±8.26 | 78.33±7.78 | 75.24±9.65 | 76.39±7.49 |
|  | MCC | LDA | 0.81±0.07 | 0.77±0.08 | 0.77±0.08 | 0.78±0.08 | 0.75±0.1 | 0.75±0.08 |
|  |  | SVM | 0.81±0.08 | 0.76±0.09 | 0.78±0.09 | 0.78±0.08 | 0.75±0.1 | 0.75±0.08 |
|  |  | KNN | 0.81±0.08 | 0.76±0.08 | 0.77±0.09 | 0.77±0.08 | 0.74±0.1 | 0.75±0.08 |

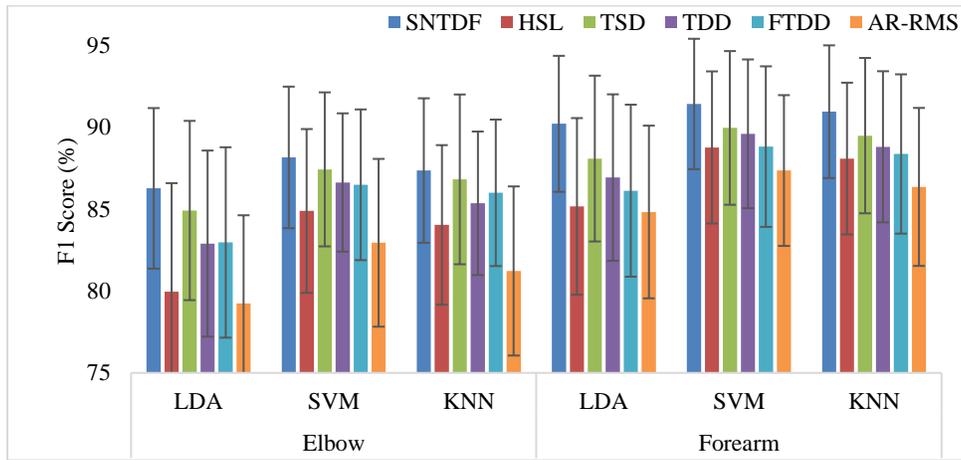
Fig. 3. Forearm orientation invariant hand gesture recognition performances training with pronation and supination.

## B. Forearm Orientation Invariant Hand Gesture Recognition Performance Training with Rest Orientation (CASE-A)

To find optimum electrode position in forearm orientation invariant hand gesture recognition, two electrode positions, six feature extraction methods, and three classifiers were considered. In forearm orientation invariant performance evaluation, three classifiers were trained with the hand gestures of rest orientation however the classifiers were tested with the hand gestures of all three orientations (pronation, rest, and supination). In this research, the other two orientations (pronation and supination) were not studied as the rest is the best orientation for providing improved forearm invariant hand gesture recognition performance [8], [34]. The average value of the performance with standard deviation is presented in Table III. As the F1 score presents true positive hand gesture recognition performance more precisely combining both sensitivity and precision, the summary of the table is also presented in Fig. 2 using the F1 score only. The experimental results indicated that the forearm orientation invariant performance is independent of the classifiers ($p=0.25$) but significantly dependent on the feature extraction methods ($p<0.05$), and electrode positions ($p<0.05$). The forearm electrode position overperformed the elbow electrode position with all performance evaluation parameters, feature extraction methods, and classifiers. In comparison among the classifiers, LDA showed a little bit higher performance than SVM and KNN. However, among the feature extraction methods, SNTDF significantly outperformed most of the feature extraction methods as the $p$-value between SNTDF and each of HSL, TSD, TDD, FTDD, and AR-RMS was <<0.001, 0.03, 0.40, 0.005, and 0.24, respectively. With best performing SNTDF and LDA, the forearm electrode position achieved the highest accuracy, sensitivity, specificity, precision, F1 score, and MCC of 97.01%, 82.05%, 98.38%, 84.17%, 81.90%, and 0.81, improving by a value of 0.93%, 5.55%, 0.51%, 4.65%, 5.35%, 0.06, respectively.

## C. Forearm Orientation Invariant Hand Gesture Recognition Training with Pronation and Supination (CASE-B)

To improve forearm orientation invariant hand gesture recognition performance, the number of training orientations was increased to two (pronation and supination). The three classifiers were trained with the hand gestures of pronation and supination orientations and tested with the hand gestures of training orientations and rest orientations as well, considering six feature extraction methods. In this study, the training combination of pronation and supination was employed as it provides improved performance than the other two combinations [31], [34]. Again, for performance evaluation, only the F1 score was considered indicating true positive hand gestures more precisely. The average forearm orientation invariant hang gesture recognition performances with their standard deviations are presented in Fig. 3. The experimental results indicated that the performance was significantly influenced by the electrode positions ($p<<0.001$) and the feature extraction methods ($p<<0.001$), however independent on the classifiers ($p=0.87$). For the best-performing SVM and SNTDF, the forearm electrode position achieved a significantly improved F1 score of 91.41% which is improved by 3.26% compared to the elbow electrode position. Also, the F1 score was higher by 9.51% compared to CASE-A. The achieved forearm orientation invariant hand gesture recognition performance was satisfactory and can be recommended for prosthetic hand or human-computer interfacing.

## D. Forearm Orientation Invariant Hand Gesture Recognition with Two Electrode Positions (CASE-C)

Forearm orientation invariant hand gesture recognition performance was also evaluated, training with the hand gestures of rest orientation of both elbow and forearm electrode positions, and tested with the hand gestures of training orientation and two unknown orientations (pronation and supination) as well. The recognition performances considering three classifiers and six feature extraction methods are presented in Fig. 4. The experimental results depicted that the classification performance depends on feature extraction methods ($p = 0.02$) rather than on the classifiers ($p = 0.69$). In this case, the highest F1 score of 87.92% was achieved with LDA and SNTDF which was improved by 6.02% compared to CASE-A. So, multiple electrode array on the forearm improves forearm orientation invariant hand gesture recognition performance. It was also noted here that the F1 score of CASE-C was less than CASE-B by 3.49%. Therefore, the number of training orientations is more effective than the number of arrays or electrode positions on the forearm for improving forearm orientation invariant hand gesture recognition.

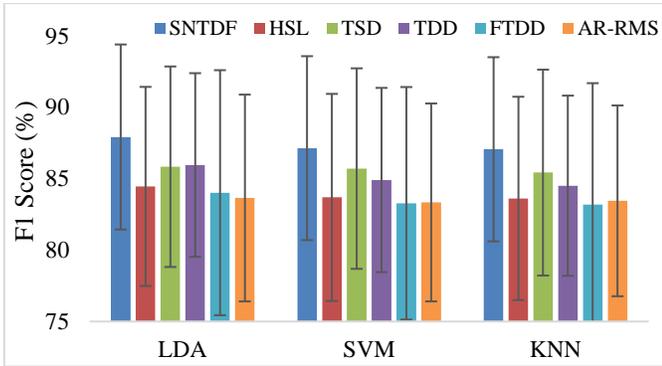

Fig. 4. Forearm orientation invariant hand gesture recognition performances with two electrode positions on the forearm.

### E. Real-time Performance

To validate experimental results obtained in CASE-A, CASE-B, and CASE-C, the real-time performances were also evaluated considering SNTDF and LDA (Table IV). In this study, four subjects from a total of 19 participated. The LDA classifier was trained with the collected data and tested with the real-time EMG signal. In CASE-A, training the classifier with rest orientation and testing with all orientations, the forearm electrode position achieved an F1 score of 71.40 % which was improved by 7.85% compared to the elbow electrode position. Again, in CASE-B, training the classifier with both pronation and supination orientations and testing with all orientations, the forearm electrode position achieved a 4.87% improved F1 score of 78.23% compared to the elbow electrode position. Therefore, the forearm electrode position consistently showed improved performance in both offline and real-time. In addition, the performance was also evaluated utilizing two electrode positions (training the classifier with rest orientation of both elbow and forearm electrode positions, and testing with all orientations) and achieved an F1 score was 81.35%. However, the real-time performance was less than the offline performance due to some unavoidable effects included in real-time scenarios, including data acquisition for a long time, a lack of proper training, unconsciousness of the subjects, etc.

TABLE IV
FOREARM ORIENTATION INVARIANT REAL-TIME PERFORMANCE

| Training Schemes | F1 Score (%) | |
|---|---|---|
| | Elbow | Forearm |
| CASE-A | 63.55±8.95 | 71.40±8.55 |
| CASE-B | 73.36±4.90 | 78.23±3.15 |
| CASE-C | 81.35±6.83 | |

### F. Performance Comparison

In this study, the forearm orientation invariant hand gesture recognition performance was compared and validated with the existing works (Table V). Khushaba *et al.* [31] evaluated the hand gesture recognition performance of both pronation and supination orientations, training with the hand gestures of rest orientation only. They observed that the average recognition performance across 12 healthy subjects dropped drastically to about 60.67%. However, they noticed about 17% higher performance of 76.77% for an amputee as known rest orientation was included in testing along with two unknown orientations (pronation and supination). Rjapriya *et al* [34] also evaluated forearm orientation invariant hand gesture recognition performance, training with rest orientation from three muscle contraction forces and testing with all three orientations from all muscle contraction forces. They achieved a higher performance of 86.43% with their proposed feature extraction method. Furthermore, Islam *et al.* [8] evaluated both forearm orientation and muscle contraction force invariant performance and achieved a performance of 78.44%. Also, they achieved a higher performance of 91.46%, training with pronation and supination orientations from medium contraction force and testing with all orientations from all contraction forces. In this study, comparable or improved forearm orientation invariant hand gesture recognition performances were achieved, considering a 4-channel only whereas previous works employed 6-channels along with a 3-D accelerometer. Also, the feature space of 34 was much lower than 39 and 96 [31], [34]. Again, a large number of hand gestures of 12 was considered for performance evaluation. In addition, the performance was evaluated with a large number of subjects which confirmed the consistency of the results. Therefore, comparable or improved forearm orientation invariant hand gesture recognition performance with a higher number of gestures with the lowest number of channels proved the novelty of this work.

## IV. DISCUSSION

Forearm orientation is one of the potentially useful tasks that is performed combinedly with different finger gestures. Three muscles (pronator teres, pronator quadratus, and supinator) are mainly responsible for changing forearm orientations [44]. The pronator teres and pronator quadratus perform forearm pronation, the supinator performs forearm supination whereas their inactive condition is denoted by rest orientation. During performing these orientations, multiple muscles are activated, and the EMG signal due to these muscles is superimposed with the EMG signal of different hand gestures. Consequently, the surface EMG signal collected on the skin is altered due to pronation or supination orientation [45]. In addition, during performing pronation or supination orientation, the radio and ulna rotate from their resting orientation, resulting in the attached forearm muscles shifting from the surface electrodes [46]. So, the impact of muscle shifting is also added to the altered EMG signal during performing forearm orientation. That's why forearm orientation may drastically degrade the hand gesture recognition performance and limit the prosthetic hand for commercialization [12], [47].

The elbow electrode position has been considered as it is a relatively available position for most amputees. In addition, from the anatomy of the forearm, it is noted that radio and ulna bones form a cross-structure [44], so the least displacement should be between the wrist and elbow (forearm electrode position). Therefore, to improve hand gesture recognition performance with different forearm orientations and optimize electrode position, two electrode positions, elbow and forearm, were investigated. The experimental results indicated that the strength of the EMG signal of the forearm electrode position was higher in amplitude (up to 1.60 times) compared to the EMG signal of the elbow electrode position. Also, the EMG signal quality evaluated by SNR and SMR was significantly comparable to or higher than the forearm electrode position. As the forearm electrode position provided better signal quality, therefore, there might be a chance of getting improved forearm orientation invariant hand gesture recognition performance. To confirm the prediction, further forearm orientation invariant hand gesture recognition performance was evaluated, considering three training strategies (CASE-A, CASE-B, and CASE-C). In CASE-A where the classifiers were trained with rest orientation and tested with all three orientations, the LDA classifier archived a significantly higher F1 score of 81.90% with SNTDF, improving by 5.35% compared with the elbow electrode position. The improvement was consistently validated by three classifiers and six feature extraction methods.

TABLE V
THE FOREARM ORIENTATION INVARIANT HAND GESTURE RECOGNITION PERFORMANCE COMPARISON.

| Reference | Dynamic Factor | EMG Dataset | | | Method | | | | | Performance (%) |
|---|---|---|---|---|---|---|---|---|---|---|
| | | Subjects | Channels | Gestures | Feature set | Dimension | Training | Testing | Classifier | |
| Khushaba et al. [31] | 3 forearm orientations and 3 muscle force levels | 12-healthy 1-amputee | 6 and 3-D accelerometer | 6 | TSD and accelerometer features | 39 | Rest orientation | Pronation and supination | SVM | 60.67 (healthy) |
| | | | | | | | Rest orientation from all force levels | All orientations from all force levels | SVM | 76.77 (amputee) |
| Rajapriya et al. [34] | 3 forearm orientations and 3 muscle force levels | 10-healthy | 6 | 6 | Wavelet bispectrum based 16 features | 96 | Rest orientation from all force levels | All orientations from all force levels | LDA | 86.43 |
| Islam et al. [8] | 3 forearm orientations and 3 muscle force levels | 10-healthy | 6 and 3-D accelerometer | 6 | SNTDF, 15-order autoregression coefficients, and accelerometer features | 5 to 17 | Rest orientation from a medium force level | All orientations from all force levels | KNN | 78.44 |
| | | | | | | 7 to 20 | Pronation and supination from a medium force level | All orientations from all force levels | KNN | 91.46 |
| This work | 3 forearm orientations | 19-healthy | 4 | 12 | SNTDF | 34 | Rest orientation | All orientations | LDA | 81.90 |
| | | | | | | | Pronation and supination orientations | All orientations | SVM | 91.41 |
| | | | | | | | Rest with both elbow and forearm electrode positions (8-channels) | All orientations | LDA | 87.92 |

The experimental results indicated that the performance was significantly dependent on the feature extraction method rather than on the classifiers. To further investigate the impact of the number of training orientations and electrode positions, the classifiers were trained with two orientations (pronation and supination) and tested with all orientations (CASE-B). Following the CASE-A, the forearm electrode position achieved a 3.26% improved F1 score of 91.41% compared to the elbow electrode position. The CASE-B significantly improved the F1 score by 9.51%, so, the number of training orientations can be recommended for getting improved forearm orientation invariant hand gesture recognition. In case of less amount of amputation or intact limbed subject, both elbow and forearm electrode positions might be available. So, both electrode positions with rest orientation were also employed for training (CASE-C). Combined electrode positions achieved a significantly improved F1 score of 87.92% improved by 6.02% compared to CASE-A. Training with multiple electrodes may be facilitated with spatial information that might be a reason behind getting improved performance with multiple electrode positions. However, in comparison between the multiple orientations and electrode positions, twice orientations provided a 3.49% improved F1 score that indicates the potential of multiple orientations rather than considering multiple electrode positions. Following the forearm electrode position, the elbow electrode position also provided improved performance training with multiple orientations, therefore, in the case of high amputation where the stump length is very low, multiple forearm orientations can be recommended for training to achieve a higher performance with the variation of forearm orientation.

This study is conducted across a large number of subjects 19 whereas the recent works include up to 12 subjects [31]. Therefore, the experimental results achieved in this study are more consistent and reliable. To the best of our knowledge, only a few pieces of recent work are found. Among them, this work provides comparable to or higher forearm orientation invariant hand gesture recognition performance, recognizing 12 hand gestures where existing work recognizes 6 hand gestures only [8], [34], [31]. In addition, this work employs the least number of electrodes and dimensionality (except Islam *et al.* [8]) without compromising the recognition performance. Therefore, the employed electrode positions on the muscles can be recommended for improved performance. In this study, the forearm electrode position shows consistent results in real-time as well and proves its robustness for practical implementation.

This research incorporated intact limbed subjects only. So, the feasibility of the forearm electrode position can also be studied for amputees of various stump lengths. In addition, this study includes the most useful two electrode positions (elbow and forearm) only whereas in a human-computer interfacing, the user may be an intact limbed subject where wrist position is also available recommended by Botros *et al.*[35]. In future studies, three-electrode positions (wrist, forearm, and elbow) can be incorporated to investigate forearm orientation invariant hand gesture recognition performance. High-density electrodes can also be recommended for investigation. In addition, multiple performance limiting factors of myoelectric pattern recognition can be incorporated to find the feasibility of these electrode positions. In this study, the performance is evaluated and validated using the three most popular classifiers. So, future studies can be performed using deep-learning algorithms.

## V. Conclusion

The impact of elbow and forearm electrode positions is investigated for forearm orientation invariant hand gesture recognition. The experimental results indicate that the forearm electrode position provides comparable to or better signal quality and significantly higher forearm invariant hand gesture recognition, improving F1 score up to 5.35% compared to the elbow electrode position. The achieved performance is comparable to or higher, considering the highest number of hand gestures and subjects with the least number of channels, and feature dimensions. The robustness of the forearm electrode position is also proved in real-time experiments that predict its practical implementation. Therefore, the forearm electrode position can be recommended over the elbow electrode position to minimize forearm invariant hand gesture recognition problems for applications in myoelectric prostheses and human-computer interaction

## Acknowledgment

The authors would like to show their sincere gratitude to the subjects for their volunteer participation.